\documentclass{ws-procs9x6}

\input{psfig}

\def\Journal#1#2#3#4{{#1} {\bf #2}, #3 (#4)}
\def\MNRAS{\em MNRAS}
\def\ApJ{\em ApJ}
\def\AJ{\em AJ}
\def\NPB{\em Nucl.~Phys.~B}
\def\Nature{\em Nature}
\def\Science{\em Science}
\def\AA{\em A\&A}
\def\AcAs{\em Acta Astron}
\def\ARAA{\em ARAA}

\def\msun{{M_\odot}}
\def\pc{\,{\rm pc}}
\def\kpc{\,{\rm kpc}}
\def\mmsun{\,{\rm mM_{\odot}}}
\def\sech{\rm sech}
\def\kms{\,{\rm kms}^{-1}}
\def\solmasspc{{\msun {\rm pc}{}^{-2}}}
\def\spose#1{\hbox to 0pt{#1\hss}}
\def\lta{\mathrel{\spose{\lower 3pt\hbox{$\sim$}} \raise
2.0pt\hbox{$<$}}}
\def\gta{\mathrel{\spose{\lower 3pt\hbox{$\sim$}} \raise
2.0pt\hbox{$>$}}}

\begin{document}

\title{DOGS THAT DON'T BARK? \\
       (The Tale of Baryonic Dark Matter in Galaxies)}

\author{N.W. Evans}

\address{Institute of Astronomy, Madingley Rd, Cambridge, CB2 1ST,
England\\ E-mail: nwe@ast.cam.ac.uk}
 
%%%%%%%%%%%%%%%%%%%%%%%%%%%%%%%%%%%%%%%%%%%%%%%%%%%%%%%%%%%%%%
% You may repeat \author \address as often as necessary      %
%%%%%%%%%%%%%%%%%%%%%%%%%%%%%%%%%%%%%%%%%%%%%%%%%%%%%%%%%%%%%%

\maketitle\abstracts{This article reviews the nature and distribution
of baryonic dark matter in galaxies, with a particular emphasis on the
Milky Way. The microlensing experiments towards the Large Magellanic
Clouds, the Andromeda Galaxy and the bulge provide evidence on the
characteristic mass and abundance of baryonic dark matter, as do
direct searches for local counterparts of dark halo populations.}

\section{Introduction}

Fifteen years or so ago, it was commonly argued; ``If we want to
believe the observations rather than our prejudice, we should take as
our best bet that dark haloes are baryonic.''~\cite{dlb} Such a
viewpoint is not often heard today.  This change-of-mind has been
enforced upon us largely by the microlensing experiments.  Particle
dark matter differs from (most types of) baryonic dark matter in that
it does not produce microlensing events.  The familiar parade of
baryonic candidates has now been whittled down, and perhaps only one
remains as a possible substantial contributor to the dark matter in
the Galaxy's halo. This review assesses the distribution of missing
matter in the Galaxy (Section 2), the likely baryonic dark matter
suspects (Section 3), the evidence from microlensing (Section 4)
and from the halo white dwarf searches (Section 5).

\section{Missing Mass in the Galaxy}

\subsection{The Inner Parts}

It is now clear~\cite{reviews} that there is little dark matter in the
inner parts of big galaxies like the Milky Way. Here, the mass budget
is dominated by the baryons in the luminous disk and the bulge.

There are three strong pieces of evidence.  First, models of the Milky
Way in which the dark halo makes little contribution within the
central $\sim 5$ kpc are already strongly supported by simulations of
the gas flow in the Galactic bar~\cite{bar}. To reproduce the terminal
velocities of the HI gas, the bar and disk must provide almost all of
the gravity force field within the inner few kpc.  Second, bars in
galaxy models having halos of moderate or high central density all
experience strong drag from dynamical friction. The bar in the Milky
Way is able to maintain its observed high pattern speed only if the
halo has a central density low enough for the disk to provide most of
the central attraction in the inner Galaxy~\cite{ds}.  Third, for the
Milky Way, there are extremely high microlensing optical depths
towards Baade's Window in the bulge.  Almost all the matter in the
inner parts of the Galaxy must be capable of causing microlensing (and
hence probably baryonic)~\cite{be}.  The central $\sim 5$ kpc of the
Milky Way contain little particle dark matter.

\subsection{The Outer Parts}

\begin{figure}[t]
\psfig{figure=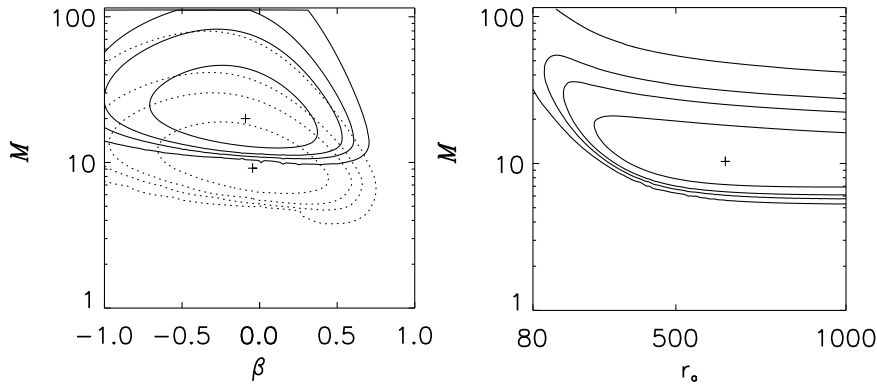,height=5.cm}
\caption{Left: Likelihood contours for the total mass $M$ of the Milky
Way halo (in units of $10^{11} \msun$) and the velocity anisotropy
$\beta$. Results including (solid curves) and excluding (dotted
curves) Leo I are shown. Contours are at heights of 0.32, 0.1, 0.045
and 0.01 of peak height and the most likely values are indicated by
plus signs. Right: Likelihood contours for the total mass $M$ of the
M31 halo (in units of $10^{11} \msun$) and the velocity anisotropy
radius $r_a$. [From Wilkinson et al. 2001]}
\end{figure}

The total mass of the Milky Way galaxy is not known very well. This is
because the gas rotation curve cannot be traced beyond $\sim 20$ kpc,
leaving only distant globular clusters and satellite galaxies as
tracers of the dark matter potential. The dataset of positions and
radial velocities (sometimes proper motions as well) of $\sim 20$
satellite galaxies and distant globular clusters is sparse. Thus, most
investigators~\cite{mw} have chosen to make strong assumptions about
the underlying halo model, using Bayesian likelihood methods to
estimate the total mass and the eccentricity of the orbits. Typical
recent results are shown in Figure 1. The solid (dotted) contours in
the left panel of Figure 1 show the likelihood including (excluding)
one of the most distant and troublesome of the satellite galaxies (Leo
I) from the dataset. The most likely total mass of the Milky Way
galaxy is $\sim 2 \times 10^{12} M_\odot$ including Leo I and $9.1
\times 10^{11} M_\odot$ excluding Leo I. For comparison, the right
panel of Figure 1 shows likelihood contours for the M31 halo, using
the dataset of projected positions and velocities of globular clusters
and satellite galaxies. The most likely mass of M31's halo is $\sim 1
\times 10^{12} M_\odot$. Given the large uncertainties in the
estimates, a reasonable conclusion is that both the Milky Way galaxy
and M31 have equally massive dark haloes. The total mass in dark
matter is about ten times greater than the total mass in stars.  The
outer parts of both the Milky Way galaxy and M31 are overwhelmingly
dominated by dark matter.

\subsection{The Solar Neighbourhood}

For all direct detection experiments, the crucial question is: how
much dark matter is there in the solar neighbourhood?  By analyzing
the line-of-sight velocities and distances of K dwarf stars seen
towards the south Galactic pole, Kuijken \& Gilmore~\cite{kg} showed
that at the solar radius there is $\sim71\pm6\,\msun\pc^{-2}$ of
material within $1.1\kpc$ of the Galactic plane. Measurements of the
proper motions and parallaxes of stars that lie within $200\pc$ of the
Sun have yielded estimates of the local density of all
matter~\cite{cphf}. For example, Cr\'ez\'e et al. found
$(76\pm15)\mmsun\pc^{-2}$ ; Pham found $(111\pm10)\mmsun\pc^{-2}$;
Holmberg \& Flynn found $(102\pm6)\mmsun\pc^{-2}$.

By counting disk M dwarfs in {\it Hubble Space Telescope} fields, the
vertical profile of these objects is known to be well modelled
by~\cite{zheng}
 \begin{equation}\label{GBFprof}
\nu(z)=0.435\sech^2(z/270\pc)+0.565\exp(-|z|/440\pc).
\end{equation}
The effective thickness of the disk's stellar mass is~\cite{be}
\begin{equation}
\hat z\equiv{1\over\nu(0)}\int_{-1.1\kpc}^{1.1\kpc}d z\,\nu(z)=691\pc.
\end{equation}
By counting stars within $5\pc$ of the Sun (which can be detected
through their large proper motions) and using Hipparcos parallaxes
Jahrei{\ss} \& Wielen~\cite{jw} found that stars contribute
$39\mmsun\pc^{-3}$ to the mass density at the plane. Multiplying this
density by the effective disk thickness $\hat z$, we have that stars
contribute $26.9\msun\pc^{-2}$ to the $71\pm6\msun\pc^{-2}$ of matter
that lies within $1.1\kpc$ of the plane. Gas (primarily hydrogen and
helium) contributes $13.7\msun\pc^{-2}$.  Thus, $\sim41\msun\pc^{-2}$
of the mass within $1.1\kpc$ of the plane can be accounted for by
stars and gas, and the remaining $\sim30\msun\pc^{-2}$ should be
contributed by dark matter~\cite{be}. The overall error on this
last number could easily be as large as $15\msun\pc^{-2}$ each way.

\section{The Usual Suspects}

This Section lines up the baryonic dark matter suspects, which could
make up some of the copious amounts of missing matter in the Galaxy.
 
\subsection{Red and Brown Dwarfs}

Red dwarfs (M dwarfs) have masses between $0.5 \msun$ and $0.08
\msun$. They shine due to hydrogen burning in their cores. Judging
from local samples, red dwarfs are about 4 times more common than all
other stars combined. About $80 \%$ of all the stars in the solar
neighbourhood are red dwarfs~\cite{reidhawley}. The local number
density of red dwarfs~\cite{rhg} as reckoned from surveys such as the
8-parsec sample is $\sim 0.07$ per cubic pc.

Brown dwarfs are objects lighter than $\sim 0.08 \msun$. They are too
light to ignite hydrogen. They are brightest when born and then
continuously cool and dim. Since 1997, near-infrared surveys (DENIS
and 2MASS) have been steadfastly uncovering brown
dwarfs~\cite{surveys}. There are over $\sim 100$ good candidates now
(as well as two new spectral classes, L and T dwarfs). The local
number density of brown dwarfs~\cite{reidetal} is very uncertain but
it may be as high as $0.1$ per cubic pc. In which case, the total
number of brown dwarfs may exceed the total number of stars in the
Galaxy.

Both red and brown dwarfs are seemingly very common in the Galactic
disk (and probably the bulge and spheroid too). But, it is now clear
that most of the missing mass in the Galactic halo cannot be ascribed
to either red or brown dwarfs. 

Red dwarfs are ruled out because they are not seen in sufficient
abundance in long exposures of high Galactic latitude fields using the
{\it Hubble Space Telescope} Wide Field Camera. More specifically,
less than 1\% of the mass of the halo can be in the form of red
dwarfs~\cite{graff}.  Brown dwarfs are ruled out because they produce
microlensing events towards the Magellanic Clouds with typical
timescales $\sim 15$ days. This is much shorter than the timescales of
the observed events, which are $\sim 40$ days. Let us recollect that
the only parameter in a microlensing event providing any physical
information is the timescale. This encodes the mass with the
velocities and distances of both the source star and the
microlens. Hence, the masses of the microlenses cannot be deduced on
an event by event basis, but typical masses can be deduced using
models of the Galactic halo.  To minimise the mass, the transverse
velocity of the microlens with respect to the line of sight must be
reduced. In the outer halo, radial anisotropy is best for doing this;
closer to the Solar circle, tangential anisotropy is best. By using a
constraint on the total kinetic energy of the lensing population, the
microlens mass can be minimised over all orientations of the velocity
dispersion tensor.  This minimum mass is $\gta 0.1 \msun$, which lies
above the hydrogen-burning limit~\cite{geg}. So, the microlenses
cannot be brown dwarfs.

\subsection{White and Beige Dwarfs}

White dwarfs are objects with mass $\sim 0.5 \msun$, the remnants of
stars with masses in the range 1-8 $\msun$. The local number density
of white dwarfs~\cite{hos} is $0.005$ per cubic pc. This is reckoned
from samples believed complete to 13 pc. If so, then white dwarfs are
about 100 times rarer than red dwarfs and brown dwarfs.

For many years, white dwarfs were regarded as very improbable
candidates for the dark matter in galactic haloes. The main problem is
that the progenitor stars are like filthy furnaces, disgorging metals
into the ISM. Carbon, nitrogen, helium and deuterium are seriously
overproduced, as judged by the present abundances of stars in the
Galactic halo~\cite{gm}. Even if all the ejecta of a population of
white dwarfs are removed by Galactic winds, the mass budget is
enormous, exceeding that of the entire Local Group. It needs a
contrived IMF so as to avoid leaving large numbers of visible main
sequence precursors still burning today in the halo. These
problems~\cite{freese} remain largely unsolved. But, the microlensing
results (with their preferred typical mass of the microlenses of $\sim
0.5 \msun$) have sparked a lot of activity in the area of white dwarf
searches -- without success so far.

Beige dwarfs have masses up to $\sim 0.2 \msun$.  These objects are
supposed to form by slow accretion of gas onto planets or brown
dwarfs. Provided the accretion energy is radiated away, the
temperature in the core never rises high enough to ignite
hydrogen~\cite{hansen}. As beige dwarfs are envisaged as primordial
objects rather than the end-points of stellar evolution, this
ingeniously circumvents the problem of pollution by metals.
Unhappily, the most recent calculations suggest that the accretion
rate needs to be $\sim 0. 1 \msun$ Gyr${}^{-1}$ -- too slow to allow
their manufacture in this Universe.

\subsection{Neutron Stars and Black Holes}

Neutron stars are the remnants of stars with initial masses in the
range 8-20 $\msun$, while black holes are the remnants of stars larger
than 20 $\msun$.  However, neutron stars and black hole remnants less
than $\sim 10^5 \msun$ cannot make up the bulk of the dark matter as
their precursors generate unacceptable metal production or background
light~\cite{carr}.

Stars larger than $\sim 10^5 \msun$ collapse directly to black holes
without excessive nucleosynthetic or background light production.
They cause microlensing events with timescales $\gta 50$ yr, which are
too long to be detectable by current surveys.  There are some
noticeable dynamical effects. For example, stellar encounters with
such black holes produce a power-law tail in the energy
distribution. Accordingly, Lacey \& Ostriker's original
paper~\cite{lo} correctly predicted the existence of the (then
unknown) thick disk.  Hence, supermassive black holes remain genuine
suspects.

\section{The Evidence from Microlensing}

Microlensing towards the Large Magellanic Cloud and the Andromeda
galaxy provides direct evidence on the fraction of dark matter in
haloes that is baryonic. Microlensing towards the bulge provide
indirect evidence on the structure of the Galactic dark halo.

\subsection{Microlensing towards the Magellanic Clouds}

The original motivation~\cite{bohd} of the microlensing experiments
was to detect the effects of baryonic dark objects in the Galactic
halo on background stars in the nearby satellite galaxies, the Large
and Small Magellanic Cloud.  From 5.7 years of data, the MACHO
collaboration~\cite{al2000} found between 13 to 17 microlensing events
towards the Large Magellanic Cloud (LMC) and reckoned that the optical
depth (or probability of microlensing) was $\tau \sim
1.2^{+0.4}_{-0.3} \times 10^{-7}$. They argued that, interpreted as a
dark halo population, the most likely mass of the microlenses is
between 0.15 and 0.9 $\msun$, seemingly implicating white dwarfs.  The
total mass in the objects out to 50 kpc is $\sim 9^{+4}_{-3} \times
10^{10} \msun$. This is $\lta 20 \%$ of the mass of the halo. In stark
contrast, after 8 years of monitoring the Magellanic Clouds, the EROS
collaboration~\cite{lasserre} secured just a ``meagre crop of three
microlensing candidates towards the LMC''.  EROS monitor a wider solid
angle of less crowded fields in the LMC. So, blending and
contamination by lenses in the LMC itself (so-called ``self-lensing'')
are much less important. EROS do not report their results in terms of
optical depth, but their experiment seemingly implies a lower value
than that preferred by MACHO.

One possibility is that the lenses lie in or close to the Large
Magellanic Cloud or in some intervening population, rather than being
true denizens of the dark halo. A number of ingenious
suggestions~\cite{noofs} have been made -- the LMC disk, tidal debris,
the warped Milky Way disk, an intervening satellite galaxy.  For one
reason or another, none of these ideas have gained a concensus,
although some contain ingredients of merit.  What we do know for sure
is that some of the lenses do not lie in the dark halo.  There are now
four exotic events~\cite{exotic} (two binary caustic crossing events,
one long timescale event with no detectable parallax, one xallarap
event) for which the location of the event can be more or less
inferred. In all four cases, the lens almost certainly lies in the
Magellanic Clouds.  Most recently of all, there has been the direct
imaging of one the microlenses by Alcock et al.~\cite{al2001},
revealing it to be a nearby low-mass star in the disk of the Milky
Way. At first glance, all this seems strong evidence that most of the
lenses do not lie in the dark halo; however, there are biases in
exotic events that favour the discovery of events in which the lens
and source are close together. In other words, {\it there is no
compelling evidence either for or against a Galactic halo origin of
the microlenses. They may equally well lie in the dark halo or they
may lie in the Magellanic Clouds or in intervening populations.}

A final possibility that deserves serious consideration is that some
microlensing events may have been misidentified. For example, there
are expected to be $\sim 20$ supernovae in background galaxies behind
the LMC and brighter than MACHO's limiting magnitude during the
experiment's lifetime; this number may be larger by at least a factor
of two, depending on the supernova contribution from faint galaxies.
So, supernova contamination is a serious problem~\cite{al2000}. The
bumps do no repeat and rise up from a flat baseline. They differ from
microlensing curves in that they are asymmetric, but such asymmetry
may not be obvious in noisy or sparsely sampled data. In fact, MACHO's
data are taken at a site where the median seeing is $\sim 2.1$ arcsec
so the quality of the data is sometimes poor. An interesting recent
breakthrough by Belokurov and co-workers~\cite{bel} has been the
development of neural networks to identify microlensing events in
massive variability surveys. This replaces judgements made by human
experts with judgements based on strict statistical criteria.  This
technique has thus far been applied only towards the Galactic bulge,
but it already hints that some events may have been misclassified. An
analysis of the events towards the LMC can be expected from this group
soon.

\begin{table}[ht]
\tbl{Parameters for the 4 POINT-AGAPE candidates. Here, $\Delta R$ is
the magnitude (Johnson/Cousins) of the maximum source flux variation,
$t_E$ is the Einstein timescale, $t_{1/2}$ is the full-width
half-maximum and $A_{\rm max}$ is the maximum amplification. All these
events have very high amplification and short full-width half-maximum
timescale. [From Paulin-Henriksson et al. 2002] .\vspace*{1pt}}
{\footnotesize
\begin{tabular}{|l|c|c|c|c|} \hline
reference & $\Delta R$ (mags) & $t_{1/2}$ (days) & $t_E$ (days)
& $A_{\rm max}$ \\ 
\null & \null & \null & \null & \null \\
PA-99-N1  & $20.8\pm 0.1$ & 1.9    & $9.74\pm 0.70$  
& $17.54^{+1.33}_{-1.15}$ \\
PA-99-N2  & $19.0\pm 0.2$ & 25.0   & $91.91^{+4.18}_{-3.83}$ & 
$13.33^{+0.75}_{-0.67}$ \\
PA-00-S3  & $18.8\pm 0.2$ & 2.3    & $12.56^{+4.53}_{-3.23}$ & 
$18.88^{+8.15}_{-5.89}$ \\
PA-00-S4  & $20.7\pm 0.2$ & 2.1    & $128.58 ^{+142.61}_{-72.27}$
& $211^{+16456}_{-120}$ \\ 
\null & \null & \null & \null & \null \\ \hline
\end{tabular}}
\end{table}
\begin{figure}[t]
\centerline{\psfig{figure=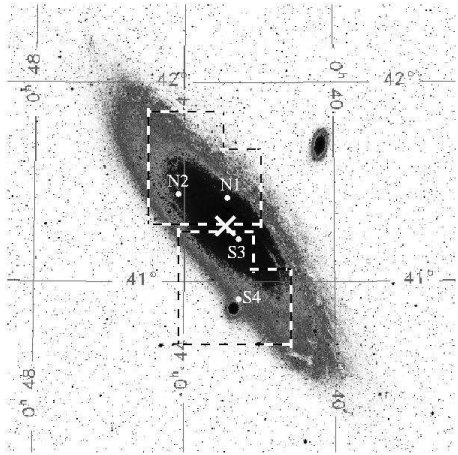,height=6.cm}}
\caption{The location of 4 microlensing events detected by POINT-AGAPE
towards M31. Also marked are the two fields that straddle the north
and south of M31. [From Paulin-Henriksson et al. 2002]}
\label{figure:pa}
\end{figure}

\subsection{Microlensing towards M31}

Microlensing experiments towards M31 have the potential to clarify the
ambiguous results towards the LMC. This is because M31 is highly
inclined ($i \sim 77^\circ$). Lines of sight to disk stars in the
north or near side of M31 are shorter than those to the south or far
side. Microlensing by a spheroidal dark halo will have a
characteristic signature with an excess of events on the far side of
the M31 disk~\cite{crotts}. This signal is absent when the microlenses
lie in the stellar disk or bulge of M31.  A number of
groups~\cite{many} are now carrying out large-scale surveys of M31 to
look for this near-far disk asymmetry.  In M31, the individual stars
are not resolved, so that the flux on the detector elements (pixels or
superpixels) is monitored. Novel techniques have been developed to
monitor flux changes of unresolved stars in the face of seeing
variations~\cite{ansari}.

Recently, the POINT-AGAPE collaboration~\cite{phone,phtwo} has
reported results from two years of data taken with the Wide Field
Camera on the 2.5m Isaac Newton Telescope. The fields are shown in
Figure 2; they are $\sim 0.3$ deg${}^2$ and located north and south of
the centre of M31. Two years of data are not yet sufficient to look
for any gradient signal, but they are enough to identify a sample of
some convincing high signal-to-noise candidates.  Table 1 lists the
characteristics of four such events, designated PA-99-N1, PA-99-N2,
PA-00-S1 and PA-00-S4. Here, N (or S) tells us whether the event
occurs in the northern or southern field, while 99 or (00) tells us
whether the event peaks in 1999 (or 2000).  The events are selected on
the basis of a set of severe selection criteria. There are $\sim 350$
candidates with a single, substantial, symmetric bump which is a good
fit to the standard Paczy\'nski form. Many of these are variable stars
and a longer baseline is needed to provide
discrimination. Accordingly, Paulin-Henriksson et al. insist upon a
short full-width half-maximum timescale ($t_{1/2} < 25$ days) and a
flux variation exceeding the flux of a 21st magnitude star ($\Delta R
<21$). The rationale for this is that microlensing is the only
astrophysical process that can cause such huge fluctuations on such
very short timescales. This cut leaves eight microlensing
candidates~\cite{phthree}, of which the four listed in Table 1 are the
most convincing.

These early results are tantalizing. Microlensing events in the inner
few arcminutes are overwhelmingly due to stellar lenses in the
bulge~\cite{kerins}. Hence, PA-99-N1 and PA-00-S3 are likely to be
caused by low mass stars in the bulge.  PA-00-S4 lies about $22'$ from
the centre of M31, but it is only $3'$ from the centre of the
foreground dwarf elliptical galaxy M32. Although the lens could be a
dark object in M31's halo, the closeness to M32 suggests that a
stellar lens in M32 is more probable~\cite{phone}. The fourth event
PA-99-N2 lies far out in the M31 disk and there are no obvious
concentrations of stellar lenses along the line of sight. At first
glance, this looks a good candidate for a lens in the dark halo of
either of our Galaxy or M31. However, the self-lensing optical depth
(that is, the probability that an M31 disk star is lensed by another
M31 disk star) is~\cite{phtwo}
\begin{equation}
\tau_{\rm disk} \approx {4\pi G\Sigma_{\rm disk} h\sec^2 i\over c^2}
\approx 2.5\times 10^{-7}\,{\Sigma_{\rm disk}\over 100\,\solmasspc}\,
{h\over 200\,\pc},
\label{eqn:diskselflens}
\end{equation}
where $\Sigma_{\rm disk}$ is the disk column density, $h$ is its
exponential scale height and we have normalised the formula to likely
values.  In fact, $\tau_{\rm disk}$ is comparable to the halo optical
depth at this location for a $20\%$ baryon fraction.  However, the
disk self-lensing hypothesis makes a reasonably model-independent
prediction about the event timescales, namely
\begin{equation}
\langle t_{\rm E}\rangle \approx{1\over c}\sqrt{{16\over\pi^3}
\,{M\over \Sigma_{\rm disk}}} \approx
100\,{\rm days}\, \biggl({ M\over 0.5\,M_\odot}\biggr)^{1/2}\,
\biggl({\Sigma_{\rm disk}\over 100\,\solmasspc}\biggr)^{-1/2}
\end{equation}
which is in good agreement with the $t_{\rm E}$ of $\sim 90$ days for
PA-99-N2.

These arguments are teasingly suggestive rather than
conclusive. Certainly, the notion that dark halo lenses are
responsible for all of these events is not especially favoured,
although it is cannot be rejected right now~\cite{phone}. In the cases
where halo lenses may be responsible (PA-99-N2, PA-00-S4 and perhaps
PA-99-N1), stellar lensing is equally likely.  Had one or several of
these events been projected against the far side of the M31 disk, well
away from the M31 bulge and from M32, then halo lensing would have
been the likely culprit.

\subsection{Microlensing towards the Bulge}

Lines of sight towards the bulge do not probe the halo dark matter
directly, as almost all the lensing events are probably caused by low
mass stars in the disk and the bulge. Rather surprisingly, however, we
do learn something concerning the structure of the Galaxy's dark halo
from these experiments.

Table~2 shows the measurements of the optical depth to microlensing to
the red clumps stars in the bulge~\cite{data}.  Red clump stars are
bright stars that are known to reside in the bulge.  The experiments
have remained very consistent with an optical depth of $\tau \gta 3.0
\times 10^{-6}$.  Figure 3 shows contours of optical depth in three
barred models of the inner Galaxy~\cite{beleva}. All three models have
been derived from the same dataset, namely the infrared surface
photometry measured by the DIRBE instrument on the COBE satellite, but
made different corrections for absorption and emission by dust. Models
such as those based on constant mass-to-light deprojections of the
infrared photometry~\cite{bgs} are not able to reproduce these high
optical depths. Freudenreich's model~\cite{freud} does come closer (as
Figure 3 shows), although it too has some difficulties with the
highest values, such as the most recent results from Sumi et al. (the
MOA collaboration).

\begin{table}[ht]
\tbl{The microlensing optical depth recorded by various experimental
groups towards locations in the Galactic bulge.\vspace*{1pt}}
{\footnotesize
\begin{tabular}{|l|c|c|}
\hline
Collaboration  & Location & Optical Depth \\
\null & \null & \null \\
Udalski et al. (1994)  & Baade's Window   & $\sim 3.3 \times 10^{-6}$ \\
Alcock et al. (1995) & ($2.3^\circ,-2.65^\circ$) & $\sim 3.25 \times
10^{-6}$ \\
Alcock et al. (1997) & ($2.5^\circ,-3.64^\circ$) & $ 3.9^{+1.8}_{-1.2}\times
10^{-6}$ \\
Alcock et al. (2000) & ($2.68^\circ,-3.35^\circ$) &
$3.23^{+0.52}_{-0.50} \times 10^{-6}$\\
Popowski et al. (2000) & ($3.9^\circ,-3.8^\circ$) & $2.0\pm 0.4 \times 10^{-6}$\\
Sumi et al. (2002)   & ($3.0^\circ,-3.8^\circ$) & $3.40^{+0.94}_{-0.73}
\times 10^{-6}$ \\ 
\null & \null & \null \\
\hline
\end{tabular}}
\end{table}
\begin{figure}[ht]
\psfig{figure=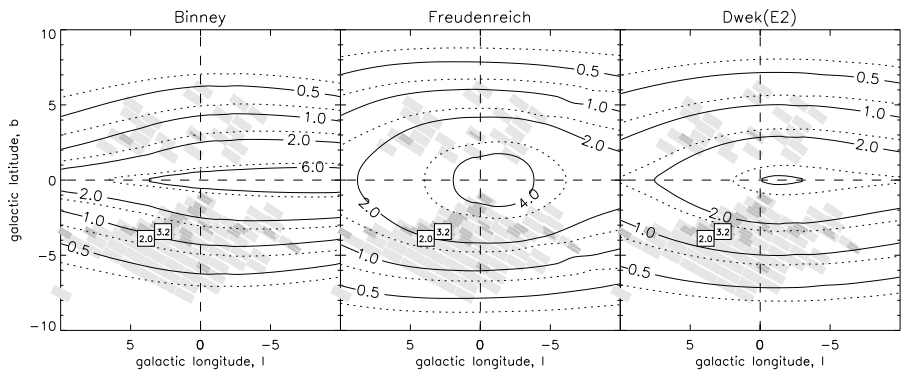,height=4.cm}
\vskip 0.5truecm
\caption{Contours of microlensing optical depth to the red clump
giants (in units of $10^{-6}$) in the three Galaxy models, excluding
(full lines) and including (dotted lines) spirality. The optical
depths reported by Alcock et al. (2000) and Popowski et al. (2000) are
shown in boxes. Light (or dark) gray boxes correspond to EROS (or
OGLE) fields. [From Evans \& Belokurov 2002]}
\end{figure}

A crucial difference between baryonic and particle matter is that the
former can cause microlensing events, while the latter cannot. To get
these high optical depths, almost all the matter permitted by the
rotation curve must be baryonic within the inner $\sim 5$ kpc. Figure
4 shows a fit to the tangent-velocity data (short dashed line)
originally derived by Binney et al.~\cite{binneyetal}. Any model must
lie below this curve. The dotted curve shows the contribution to the
rotation curve from a bulge and disk judged to reproduce an optical
depth of $2 \times 10^{-6}$ towards Baade's Window (itself a
conservative value, lower than most of the measurements in Table
2). The long-dashed curve shows the contribution of the dark halo to
the rotation curve using the local column densities of dark matter
derived in Section 2.3, assuming the cusped Navarro-Frenk-White (NFW)
model~\cite{nfw} currently favoured by cosmological simulations. The
total rotation curve always lies above the data. {\it The high optical
depths to microlensing seen towards the bulge are enough to rule out
cusped dark halo models like Navarro-Frenk-White for the Milky
Way~\cite{be}.}

\begin{figure}[t]
\psfig{figure=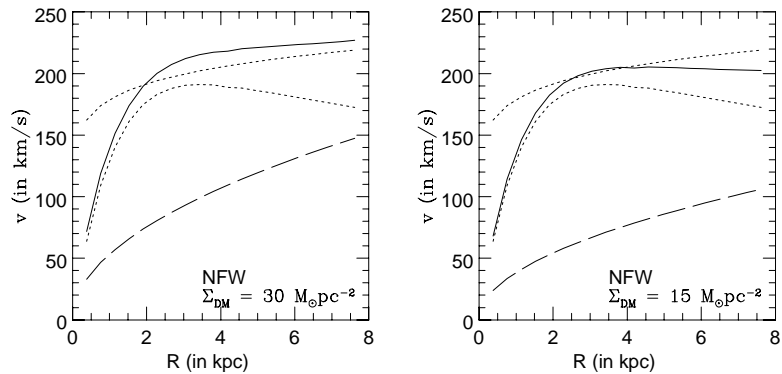,height=6.cm}
\caption{The panels show the circular-speed curves generated by the
gas disk together with enough stars to yield $\tau = 2 \times 10^{-6}$
(dotted curves) and by NFW haloes (long-dashed curves). The combined
rotation curve is shown as a solid line and must lie below the fit to
the tangent-velocity data (short-dashed line). The panels differ in
the local dark matter column density [From Binney \& Evans 2001].}
\end{figure}

\section{The Evidence from White Dwarf Surveys}

There have been a number of searches for local examples of high
velocity, very cool white dwarfs that might be representatives of the
halo population causing the microlensing towards the LMC.

First, Ibata et al.~\cite{ibata} claimed the detection of 5 faint
objects with significant proper motion in the Hubble Deep Field. They
argued that the observations were consistent with old white dwarfs
with hydrogen atmospheres. They claimed that this provided a local
mass density which, if extrapolated, was sufficient to account for the
microlensing results.  Strictly speaking, Ibata et al. found 5 faint
objects whose light centroids shifted between the first and second
epoch of exposures (separated by 2 years). For point sources, such
centroid shifts might be indicative of proper motions; however, they
also can arise easily enough for extended or variable sources.
Richer~\cite{richer} withdrew the results after an analysis of the
third epoch data, taken 5 years after the original Hubble Deep
Field. None of the 5 objects possessed a statistically significant
proper motion. Whatever the objects are, they are not moving and so
certainly not high velocity white dwarfs.

Second, Oppenheimer et al.~\cite{opp} also claimed ``direct detection
of galactic halo dark matter''. They identified candidates with
sub-luminosity and with high intrinsic proper motions from SuperCOSMOS
Sky Survey Plates. They followed this up with spectroscopy to discover
38 cool white dwarfs.  Oppenheimer et al. derive $U$ and $V$
velocities for each target by setting $W=0$, where ($U,V,W$) are the
components towards the Galactic Centre, in the direction of Galactic
rotation and perpendicular to the Galactic plane respectively. Systems
with $[U^2 + (V+35)^2]^{1/2} > 94 \kms$ are identified as members of
the halo. They derive a halo white dwarf density of $\gta 2.2 \times
10^{-4}$ stars per cubic pc. This is a factor of 10 higher than the
expected density of white dwarfs in the stellar halo. However, this
result has been contested by Reid et al.~\cite{reid}, who question the
validity of the kinematic discriminant. For example, there is a
significant excess of prograde rotators in Oppenheimer et al.'s
sample: 34 out of 38.  This is exactly the behaviour expected if a
substantial number of the white dwarfs are drawn from the thick disk
rather than the halo. Most true halo populations are only weakly
rotating.

Despite false alarms, such halo white dwarf surveys are clearly worth
pursuing.  The discovery of a local counterpart to the putative
microlensing population would be a substantial breakthrough.

\section{Conclusions}

Evidence from dynamics and particularly microlensing has made many
baryonic dark matter candidates unlikely as components of galaxy
haloes.  The constraints on stellar baryonic dark matter are
especially harsh, with brown, red, beige and white dwarfs ruled out as
dominant contributors. Only at the very high mass end (supermassive
black holes) do possibilities remain for building the Galactic halo
entirely from dark baryonic objects.

It is curious that none of the microlensing events towards the
Magellanic Clouds or Andromeda can be ascribed to lenses in the dark
halo with surety. Some of the events have been almost certainly
identified with stellar populations. This includes the exotic lenses
towards the Magellanic Clouds and some of the events towards M31.
{\it This need not have been the case}. Unambiguous halo candidates
could have been found -- for example, binary caustic crossing events
implicating a halo lens in the experiments towards the Magellanic
Clouds or short timescale events far out in the M31 disk. Similarly,
local searches could have identified a convincing counterpart to any
halo baryonic dark matter population -- but did not!

The dogs could have barked three times in the night~\cite{conandoyle}
(during the MACHO experiment, in the POINT-AGAPE datasets, in the
white dwarf searches). Each time, the dogs stayed silent.

\end{document}